\documentclass[floatfix,aps,twocolumn,pre,superscriptaddress]{revtex4-1}
\usepackage{graphicx}
\usepackage{dcolumn}
\usepackage{bm}
\usepackage[T1]{fontenc}
\usepackage{textcomp}
\usepackage{amsmath}
\usepackage{mathrsfs}
\usepackage{graphicx}
\usepackage{amssymb}
\usepackage{units}
\usepackage{xcolor}
\usepackage{float}

\newcommand{\iac}[1]{#1}

\begin{document}

\title{Kardar-Parisi-Zhang universality in the linewidth of non-equilibrium 1D quasi-condensates}

\author{Ivan Amelio}
\affiliation{Center for Nonlinear Phenomena and Complex Systems,
Universit{\'e} Libre de Bruxelles, CP 231, Campus Plaine, B-1050 Brussels, Belgium}
\affiliation{
Institute of Quantum Electronics ETH Zurich, CH-8093 Zurich, Switzerland}
\affiliation{Pitaevskii BEC Center, INO-CNR and Dipartimento di Fisica, Università di Trento
I-38123 Trento, Italy}
\author{Alessio Chiocchetta}
\affiliation{Institute for Theoretical Physics, University of Cologne, Zülpicher Strasse 77, 50937 Cologne, Germany}
\author{Iacopo Carusotto}
\affiliation{Pitaevskii BEC Center, INO-CNR and Dipartimento di Fisica, Università di Trento
I-38123 Trento, Italy}

\begin{abstract}
We investigate the finite-size origin of the emission linewidth of a spatially-extended, one-dimensional non-equilibrium condensate.
We show that the well-known Schawlow-Townes scaling of laser theory, possibly including the Henry broadening factor, only holds for small system sizes, while in larger systems the linewidth displays a novel scaling determined by Kardar-Parisi-Zhang physics. This is shown to lead to an opposite dependence of the linewidth on the optical nonlinearity in the two cases.  
We then study how sub-universal properties of the phase dynamics such as the higher moments of the phase-phase correlator are affected by the finite size and discuss the relation between the field coherence and the exponential of the phase-phase correlator.
We finally identify a configuration with enhanced open boundary conditions, which supports a spatially uniform steady-state and facilitates experimental studies of the linewidth scaling.
\end{abstract}

\date{\today}

\maketitle



\section{Introduction}


The statistical theory of critical phenomena in systems at thermal equilibrium is considered one of the most successful branches of theoretical physics~\cite{1963huang,2010brezin}.
The central result is that at criticality the low energy correlation functions
are characterized   by universal
exponents, insensitive to the microscopic details of the system
but only determined by  dimensionality and symmetries, in particular spontaneously broken ones.
As a result, seemingly distant phenomena such as the gas-liquid transition and Ising ferromagnetism end up belonging to the same universality class.
While the general theory refers to a spatially infinite setting, real systems necessarily have a finite spatial extent, so that the study of finite-size effects is of great importance in this context.  
In particular, finite-size scaling methods~\cite{fisher1972,cardy2012finite} have been developed to obtain precise estimates of the critical exponents from measurements on systems of different sizes on the order of the correlation length. Such a tool has turned out to be of tremendous utility in numerical simulations. 

Even though criticality is omnipresent also in out-of-equilibrium systems -- power law correlators are present for instance in avalanches, percolation, social networks, and many other natural phenomena --, here a general classification scheme is still lacking. 
In contrast to equilibrium where scale invariance only appears in the proximity of phase transitions and requires a fine-tuning of one or more parameters, in non-equilibrium systems it can be observed without a fine-tuning of the parameters, hence the name of  self-organized criticality~\cite{bak1988self}. This provides a further motivation to explore the interplay between finite-size effects and universality in non-equilibrium systems.

In this work we study the linewidth of a driven-dissipative 1D (quasi-)condensate.
Experimentally relevant platforms to investigate this physics include lasing in 1D spatially extended systems such as photonic~\cite{zhang1995} or polariton~\cite{wertz2010} wires, discrete arrays of polariton micropillars~\cite{amo2016,fontaine2022} or
VCSELs~\cite{grabherr1999,cui2009}, or even the edge modes of 2D topological lasers~\cite{harari2018,bahari2017,loirette-pelous2021,amelio2020theory}. 
In all these systems, a natural and technologically very relevant observable is the emission linewidth, namely the spectral width of the light emitted from a given point of the device.

Schawlow and Townes in their seminal work~\cite{schawlow1958} predicted that the ultimate linewidth of a single mode laser is set by the spontaneous emission and scales inversely to the number of photons in the laser cavity. 
This scaling is accurate for the simplest, textbook case of a zero-dimensional device where the spatial dynamics of the light field is frozen and a single cavity mode can be considered.
Ramping up in geometrical complexity, two of us recently remarked in~\cite{amelio2020theory} how a finite value of the linewidth of a spatially extended laser device can be viewed as a finite-size effect.
While in a zero-dimensional system the phase dynamics is diffusive and the phase-phase correlator grows linearly in time, a different behaviour is anticipated for infinite-size, low-dimensional systems, where the phase-phase correlator is characterized by specific exponents. As proven in a series of recent studies \cite{gladilin2014,altman2015,he2015,squizzato2018}, non-equilibrium 1D quasi-condensates belong in fact to the 
Kardar-Parisi-Zhang (KPZ)~\cite{kardar1986} universality class.
Here we undertake the study of finite-size effects in this context and, in particular, show that, for large enough systems, the signatures of the KPZ universal exponents are visible in the scaling of the linewidth with the spatial size of the device. This provides an exciting link between crucial concepts of non-equilibrium statistical mechanics and an observable quantity of central importance in the general theory of lasing as well as for technological applications.

The structure of this work is the following.
In Sec.~\ref{sec:KPZpolaritons} we start introducing the model for the quasi-condensate and reviewing the general KPZ theory of the phase dynamics.
In Sec.~\ref{sec:scaling}
we report numerical simulations of the stochastic Complex Ginzburg Landau Equation (CGLE) describing the field evolution and we compare the result with the predictions of the Kuramoto-Sivashynskii equation (KSE) describing the phase dynamics and of the low energy KPZ equation. 
For both the full CGLE dynamics and the KSE,  the numerical calculations clearly show that  for small system sizes $L$ the coherence time (or inverse linewidth) scales linearly in  $L$ according to the Schawlow-Townes prediction, while for large systems the scaling is instead proportional to $\sqrt{L}$. For the low-energy KPZ evolution, only the latter scaling is observed, as entailed by the universal  1D KPZ exponents under a finite-size scaling hypothesis.
The KPZ scaling properties are then used to explore the effect of an optical nonlinearity, namely a photon-photon interaction term, on the linewidth:
interestingly, this nonlinearity has opposite effects in the two regimes, so the optimal linewidth is obtained  for an intermediate value of the interactions.

In Sec.~\ref{sec:skewness} we investigate the higher moments of the 
phase-phase correlator, that are known to exhibit sub-universal features~\cite{takeuchi2018appetizer,squizzato2018}. In particular, we present numerical evidence that, as a finite-size effect, the probability distribution for the phase transits from a skewed distribution, as expected by KPZ, to an approximately Gaussian distribution at large times.
This entails that re-exponentiation of the phase-phase correlator is allowed when computing the linewidth.

In view of facilitating experiments, in Sec.~\ref{sec:boundaryconditions} we propose a lattice configuration with enhanced open boundary conditions: the spatially uniform profile of the steady-state is of great utility for the accurate extraction of the scaling of the linewidth with $L$ in experiments. 
Conclusions are finally drawn in Sec.~\ref{sec:conclusions}

\section{KPZ universality in one-dimensional non-equilibrium condensates}
\label{sec:KPZpolaritons}

We start by reviewing the theory of KPZ universality in one-dimensional geometries~\cite{gladilin2014,he2015,altman2015,squizzato2018,amelio2020theory}.
Even though we will  focus on the case of a continuous wire, the main results also apply to discrete lattices, e.g. the Lieb arrays of polariton micropillars considered in~\cite{fontaine2022} as well to the edge modes of 2D topological lasers~\cite{amelio2020theory,someth-pelous2021}.
Assuming that the reservoir of carriers can be adiabatically eliminated, the field dynamics is described by the stochastic complex Ginzburg-Landau equation (CGLE)
\begin{equation}
i \partial_t \psi =  
\left[ - \frac{1}{2m} \nabla^2
+ g n +  \frac{i}{2} \left( \frac{P}{1+ n/n_S}  - \gamma 
\right) \right]  \psi 
+ \sqrt{2D}\xi,
\label{eq:GPE}
\end{equation}
where $\psi$ is the semi-classical field, $m$ the photon mass, $g$ the strength of the Kerr optical nonlinearity, namely the photon-photon interactions, and  $n=|\psi|^2$ the density. The non-equilibrium features enter through the loss rate $\gamma$, the effective pumping rate $P$ and the saturation density scale $n_S$.
Finally,  $\xi$ is a Gaussian-distributed white noise term $\langle \xi^*(x,t) \xi(x',t') \rangle =  \delta(x-x')\delta(t-t')$ and $D$ is the  noise strength coefficient.

At mean-field level, the steady-state  above the condensation threshold $P>P_{th}=\gamma$  is characterized by the density $n_0 = n_S (P/P_{th} - 1)$.
 Because of the $U(1)$ symmetry of the CGLE, the global phase of the steady-state is spontaneously selected and the Bogoliubov excitation spectrum (reviewed in the Appendix) contains a gapless branch. 
 
Upon inclusion of noise, provided density fluctuations are small, one can focus on the phase dynamics, which occurs on much longer timescales compared the density relaxation rate
$\Gamma = \frac{\gamma(P-\gamma)}{P}$.
By adiabatically eliminating the density fluctuations, the CGLE \eqref{eq:GPE} reduces to the Kuramoto-Sivashinsky equation (KSE)~\footnote{Note how our Eq.~(\ref{eq:KSE})
differs from the equation considered in~\cite{gladilin2014} in that we are not making the near-threshold approximation 
$\alpha \ll 1$.}
\begin{equation}
\partial_{{t}} {\phi} =  \frac{1}{2m} \left[ -\frac{\Gamma^{-1}}{2m} \partial^4_{{x}
} {\phi} + 
\alpha \partial^2_{{x}} {\phi}  
- (\partial_{{x}}   {\phi} )^2 \right] + \sqrt{\frac{D(1+\alpha^2)}{n_0}} {\xi_1},
\label{eq:KSE}
\end{equation}
where we have introduced the blueshift  of the unperturbed steady-state $\mu = g n_0$ and the so-called Henry factor $\alpha = 2\mu / \Gamma$: the fluctuations of the density determine a local variation of the refractive index of the optical medium, which results in  an extra noise source in the phase equation, the so-called Henry linewidth broadening effect~\cite{henry1982}. Note that the phase $\phi$ indicates here an {\em unwound} phase variable that is not restricted to the $[0,2\pi]$ interval. As such, the theory based on the KSE \eqref{eq:KSE} does not capture the physics of (spatio-temporal) vortices discussed in~\cite{he2017,fontaine2022}: this approximation is legitimate as long as noise is sufficiently weak and density fluctuations are small. 



Measuring space, time, blueshift and (unwound) phase  in terms of the characteristic scales defined in terms of the microscopic parameters by
$$
l^* = \left[ (2m)^4 \Gamma^3 Dn_0^{-1}  \right]^{-1/7}, 
\ \ 
t^* = \left[ (2m)^2 \Gamma^5 (Dn_0^{-1})^4  \right]^{-1/7}, $$
\begin{equation}
\phi^* = \left[ \frac{2m(Dn_0^{-1})^2}{\Gamma}  \right]^{1/7} , \ \
\mu^* = \frac{1}{2}\left[ 2m\Gamma^6(Dn_0^{-1})^2  \right]^{1/7}.
 \label{eq:star_units}
\end{equation}
leads to the adimensional form of the KSE, 
\begin{equation}
\partial_{\tilde{t}} \tilde{\phi} = 
\tilde{\mu} \partial^2_{\tilde{x}} \tilde{\phi}
- \partial^4_{\tilde{x}} \tilde{\phi}  - (\partial_{\tilde{x}}   \tilde{\phi} )^2 + 
\sqrt{1+\alpha^2}
 \ \tilde{\xi}
\label{eq:KSE_tilde}
\end{equation}
which, at large distances and long times, renormalizes to a KPZ equation of the form~\cite{ueno2005}
\begin{equation}
\partial_{\tilde{t}} \tilde{\phi} =  \nu \partial^2_{\tilde{x}
} \tilde{\phi}  + \frac{\lambda}{2}(\partial_{\tilde{x}}   \tilde{\phi} )^2  + \sqrt{\mathcal{D}} {\xi_1}.
\label{eq:KPZ}
\end{equation}
In particular, the Galilean invariance of the KSE and KPZ~\cite{takeuchi2018appetizer} dictates that the coupling of the nonlinear term does not get renormalized and remains fixed to $\lambda = -2$ along the renormalization flow.

Neglecting finite-size effects, it can be shown~\cite{kardar1986} that the phase-phase  correlator (corresponding to the height-height correlator in the KPZ literature)  
\begin{equation}
\Delta\tilde{\phi}_{\tilde{x},\tilde{t}}^2 \equiv \langle [\tilde{\phi}(\tilde{x},\tilde{t}) - \tilde{\phi}(0,0)]^2  \rangle
\label{eq:phase-phase-correlator}
\end{equation}
has the scalings 
$\Delta\tilde{\phi}_{\tilde{x},0}^2 \sim \tilde{x}^{2\chi}$ and $\Delta\tilde{\phi}_{0,\tilde{t}}^2 \sim \tilde{t}^{2\chi/z}$ in space and time, respectively.
In $1$D the correlator is known exactly~\cite{prahofer2004}, giving values $\chi=1/2$ and $z=3/2$ for the so-called roughness and dynamical exponents.

From these results for the phase-phase correlation function, one may be tempted to perform an exponentiation of the phase variance to directly extract the field coherence
\begin{equation}
g^{(1)}(t) = \frac{1}{n_0} \left|  \langle \psi^*(x, t) \psi(x, 0) \rangle  \right|
\sim e^{-\frac{1}{2}
\Delta\tilde{\phi}_{\tilde{x},\tilde{t}}^2
}\,,
\label{eq:cumulant}
\end{equation}
which is the typical quantity that is measured in optical experiments~\cite{fontaine2022}. 
However, as it was remarked in~\cite{squizzato2018}, this ``cumulant approximation'' procedure is not generally legitimate, since the KPZ height profile at a given point is not a Gaussian random variable, but is rather given by
\begin{equation}
\phi(x,t) - \phi(x,0) 
= v_{\infty} t + \sigma t^{1/3} X + ...
\label{eq:phaseRV}
\end{equation}
where $X$ is a non-Gaussian random variable. For a system at the steady-state, the distribution of $X$ is of the Baik-Rains type~\cite{baik-rains2001}; under different initial or boundary conditions, the distribution may fall in other KPZ universality subclasses, see \cite{takeuchi2018appetizer} for a review. 

As we are going to see in what follows, in spite of the non-Gaussianity of $X$, the re-exponentiation encoded in Eq.~(\ref{eq:cumulant}) can still be used in the very long time regime to extract the emission linewidth.

\section{Scaling of the linewidth}
\label{sec:scaling}

\begin{figure}[t]
\centering
\includegraphics[width=0.48\textwidth]{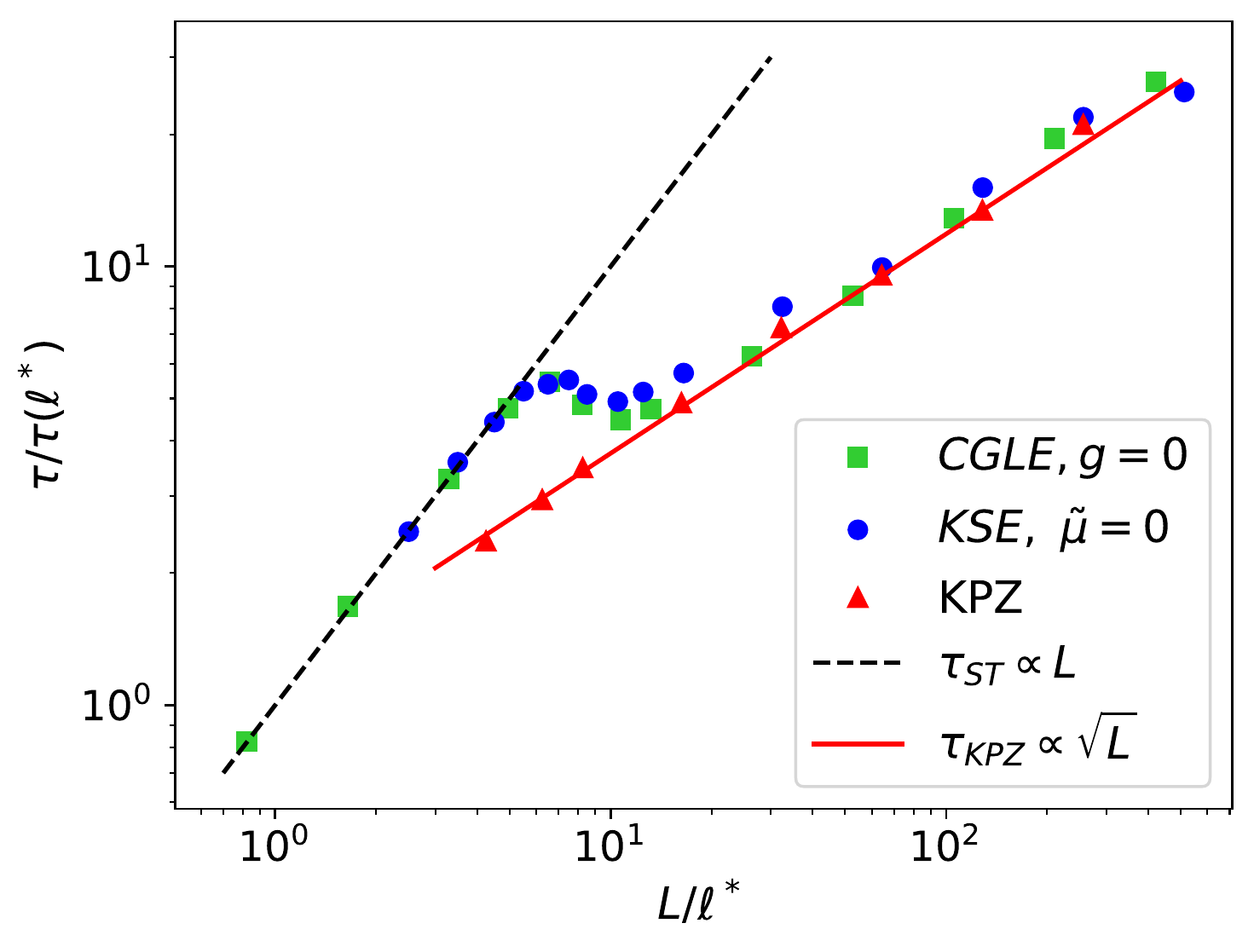}  
\caption{  Coherence time $\tau_c(L)$ of a one-dimensional condensate plotted as a function of the system length 
$L/l^{*}$
measured in natural units. The coherence time is here normalized to $\tau_c(l^{*})$ as computed for $L=l^*$.  The three numerical data series  correspond to the full CGLE  (green squares) and the KSE Eq.~(\ref{eq:KSE_tilde}) (blue circles)
and KPZ Eq.\eqref{eq:KPZ} (red triangles) approximations.
The interaction constant is set to zero, $g=0$, and the other CGLE parameters are $\gamma = 0.1$, $m=10$, $P =2\gamma$, $D=\gamma$, $n_S=1000$.
In the present $g=0$ case, the KSE  has $\tilde{\mu},\alpha = 0$. The parameters of the KPZ are chosen as $D/\nu=1.82,\lambda=2$, so  to match the same universal long-wavelength physics as the CGLE and KSE.
For small sizes Bogoliubov theory holds and the coherence time scales  proportionally to the system length as predicted by the Schawlow-Townes theory (black dashed line).
For longer systems, instead, the  nonlinearity of the phase dynamics determines a stronger broadening of the linewidth and a scaling $\tau_c \propto L^{1/2}$ (red line).
}
\label{fig:Lscaling}
\end{figure}

While these universal features have been derived for the case of spatially infinite systems, they provide an accurate description also for finite systems up to a saturation time scaling as $L^z$. For longer times/shorter systems, the physics is instead dominated by finite-size effects, which are also expected~\cite{amelio2020theory} to display remarkable features as we are now going to see.

If spatial fluctuations are neglected in a sort of single-mode approximation, the long time behavior of the correlation function is dominated by the diffusion of the phase. Since any restoring force is forbidden by the microscopic $U(1)$ symmetry of the model, the phase performs a random walk in time. In this case,
re-exponentiation \eqref{eq:cumulant} is exact  and yields an exponential decay of the coherence function 
\begin{equation}
g^{(1)}(t) =  e^{- \frac{\gamma_{ST}}{2}|t| },
\end{equation}
at a rate
\begin{equation}
\gamma_{ST} = \frac{D}{n_0 L}(1 + \alpha^2)\,.
\label{eq:Schawlow--Townes}
\end{equation}
 As it was first pointed out by Schawlow and Townes~\cite{schawlow1958}, the coherence time $\tau_c \equiv 2/\gamma_{ST}$
 is proportional to the total number $N_{ph}=n_0 L$ of photons present in the lasing mode and this scaling directly extends to generic dimensions.
In all those cases when this scaling holds, we will refer to $\gamma_{ST}$ as  the Schawlow-Townes (ST) linewidth.
The factor $(1 + \alpha^2)$ was later introduced by Henry to account for the additional broadening due to refraction index fluctuations with no change in the overall scaling~\cite{henry1982}. 

As it is discussed in detail in \cite{gladilin2014,he2015,amelio2020theory}, 
for spatially extended yet sufficiently small systems, the spatial fluctuations of the phase remain weak, so one can neglect the nonlinear term of the KSE/KPZ dynamics. As a consequence, the Bogoliubov theory of non-equilibrium condensates~\cite{wouters2007} can be used and a ST scaling holds~\cite{amelio2022bogoliubov}.
Except at very short times when density fluctuations are still important, the equal-space correlators mostly probe the ST physics of the single condensate mode.

For larger systems, instead, a crossover between KPZ physics and a Schawlow-Townes-like exponential decay of coherence is visible, the latter behaviour taking over at times longer than a characteristic saturation time proportional to $L^z$. Nonetheless,
as anticipated in \cite{amelio2020theory},
the nonlinearity of the phase equation keeps having a key impact and 
results in a stronger broadening of the linewidth compared to the standard ST prediction. We therefore coin the expression generalized Schawlow-Townes (gST) regime to indicate the long-time behaviour where $g^{(1)}(t)$ decays exponentially but the standard  ST scaling $\tau_c \propto N_{ph}$ no longer holds. In the following of this Section, we specifically investigate the dependence of the coherence time $\tau_c$ on the system size $L$ in one dimension and we highlight the possibility of different scalings.

\subsection{Non-interacting $g=0$ case}

We start from the the non-interacting case $g=0$ with parameters that are chosen in a way to have
relatively small density fluctuations of the order of 10\% but sizable spatial fluctuations of the phase.

\subsubsection{CGLE simulations}
The coherence time extracted from an exponential fitting 
$g^{(1)}(t) \sim e^{-t/\tau_c}$ of a numerical simulation of the CGLE is reported in Fig.~\ref{fig:Lscaling} as green squares.
For small systems of length up to $L \sim 5 l^*$,  Bogoliubov theory holds and the effect of the KPZ nonlinearity in the phase equation is negligible: correspondingly, the coherence time scales as the system size, as predicted by the ST formula (\ref{eq:Schawlow--Townes}).
For longer systems with $L > 10 l^*$, the behavior changes and the scaling is well captured by a different scaling law, $\tau_c \propto L^{1/2}$. 
At the crossover between the two regimes, a peculiar non-monotonic feature is visible whose understanding remains an open question.

In order to understand the $\tau_c \propto L^{1/2}$ scaling, we can put forward the following argument based on a finite-size scaling assumption. 
Since the KPZ equation is by itself scale invariant,  the only available length scale is provided by the system size. On this basis, we can expect that the equal-space correlator has the universal form
\begin{equation}
\Delta \phi_{0,t} \sim t^{2\chi/z} f\left(\frac{t}{L^{z}}\right).
\label{eq:scalingf}
\end{equation} 
where the function $f$ must asymptotically recover the KPZ result at small times, $f(x \to 0) \sim 1$, and  the ST behavior at long times,
 $f(x \to \infty) \sim x^{1 - 2\chi/z} $. 
 In particular, in this latter regime we have 
$
\Delta \phi_{0,t} \sim t/L^{z-2\chi}
$,
which implies that 
\begin{equation}
\tau_c \propto  L^{z-2\chi}.
\label{eq:scalingt}
\end{equation}
and thus recovers the numerically observed scaling $\tau_c \propto L^{1/2}$. 

In three (or higher) dimensions, long range order of the condensate is robust~\cite{Sieberer:PRL2013}, so we can expect that spatial fluctuations of the phase do not affect the long time coherence. As a result, the simple Schawlow-Townes scaling formula Eq.~\ref{eq:Schawlow--Townes} should hold. 
The physics is more subtle in the intermediate 2D case. Here, we can make use of the general KPZ result $\chi+z=2$  and the known exponent $\chi \simeq 0.373$~\cite{kloss2012} to predict  $\tau_c \sim L^{0.881} $. In view of a numerical verification of this prediction, a challenging issue will be to properly isolate the KPZ physics from competing effects related to the proliferation of vortices~\cite{zamora2017,he2017,ferrier2020}. This will be the topic of future numerical work.



\subsubsection{KSE and KPZ simulations}
Coming back to the one-dimensional case, it is important to verify that the behavior observed in Fig.~\ref{fig:Lscaling} does not arise from spurious physics related to the UV sector and/or to density fluctuations nor from
a violation of the cumulant approximation We have then directly simulated the KSE Eq.~(\ref{eq:KSE_tilde}) with $\tilde{\mu}=0$.
At long times, we observe that the phase-phase correlator~\eqref{eq:phase-phase-correlator} grows linearly in time with a coefficient defined as twice the inverse coherence time, $\Delta\tilde{\phi}_{\tilde{x},\tilde{t}}^2 \sim
2 t /\tau_c$. 

In Fig.~\ref{fig:Lscaling} we show as blue circles the coherence time predicted by the KSE for different system sizes. These points show an excellent agreement with the predictions of the full CGLE. 
This means that in the small density fluctuation regime studied here, the KSE description can be considered an accurate approximation for all system sizes and the cumulant approximation is a legitimate approximation, at least  for what concerns the emission linewidth. Quite interestingly, also the crossover behavior is very well reproduced by the KSE: this suggests that the kink is determined by the renormalization of the KSE \eqref{eq:KSE} and occurs at the emergent lengthscale at which the KSE flows into the KPZ.

Further insight on this is provided by numerical simulations of the KPZ equation \eqref{eq:KPZ}, whose predictions for the coherence time is shown as red triangles in Fig.~\ref{fig:Lscaling}. 
The parameters  $D/\nu = 1.92, \lambda=2$ in the KPZ equation were chosen to match the scaling function yielded by the CGLE and KSE for large system sizes (see Fig~6.b of \cite{amelio2020theory}). In other words, the  KSE renormalizes to a KPZ, whose parameters  can be obtained by extracting the scaling function from the numerical data. 
Contrary to the KSE, the KPZ equation does not capture the phase dynamics for small sizes, but the matching is excellent for systems that are long enough for the dynamics to be renormalized into the KPZ equation.

For the following, it is useful to obtain a precise expression of the  linewidth $\gamma_{KPZ}$ of the pure KPZ equation in terms of the parameters $\nu,\lambda,D$. To this purpose, let us recall 
that
the KPZ equation 
\begin{equation}
    \partial_t \phi =
    \nu \partial^2_{x} \phi - \frac{\lambda}{2} (\partial_{x} \phi )^2 + \sqrt{D} \xi,
\end{equation}
with 
$\langle \xi(x,t) \xi(0,0) \rangle = \frac{1}{2}\delta(x) \delta(t)$
is scale-invariant and can be rescaled through
\begin{equation}
    x = \frac{\nu^3}{\lambda^2 D} y, \ \ \
    t = \frac{\nu^5}{\lambda^4 D^2} s, \ \ \
    \phi = \frac{\nu^5}{\lambda^4 D^2} \varphi,
\end{equation}
to an adimensional form
\begin{equation}
    \partial_s \varphi =
    \partial^2_{y} \phi - \frac{1}{2} (\partial_{y} \varphi )^2 +  \eta , 
\end{equation}
with $\langle \eta(y,s) \eta(0,0) \rangle = \frac{1}{2}\delta(y) \delta(s)$ and no free parameters.

From the previous argument on the scaling function, we infer that the linewidth predicted by this equation for a system of extension $y \in [0,\mathcal{L}]$ has the functional form
\begin{equation}
\gamma_{KPZ}(\nu=1, \lambda=1, D=1, \mathcal{L}) =
    \lim_{s\to \infty} 
\frac{1}{s} \langle \Delta \varphi ^2 \rangle
= \frac{\gamma_1}{\sqrt{\mathcal{L}}}\,.
\end{equation}
This scaling is confimed by the red triangles in Fig.~\ref{fig:Lscaling}, from which we can fit the constant
$\gamma_1 = 0.172 \pm 0.005$.

In terms of the physical variables (in particular the system extension is $x \in [0,L]$) this reads
\begin{equation}
\gamma_{KPZ}(\nu, \lambda, D, L) =
    \lim_{t\to \infty} 
\frac{1}{t} \langle \Delta \phi ^2 \rangle
=
\frac{\lambda D^{3/2}}{\nu^{3/2}}
\frac{\gamma_1}{\sqrt{{L}}} ,
\label{eq:KPZ_linewidth}
\end{equation}
where $\gamma_{KPZ}$ only depends on the ratio $D/\nu$. This form is consistent with the fact that in the RG flow of the 1D KPZ the two parameters $D$ and $\nu$ separately diverge but the fixed point is determined by their ratio $D/\nu$ and by the $\lambda$ parameter, which is also not renormalized. A most interesting feature of Eq.~(\ref{eq:KPZ_linewidth}) is that it allows to predict the linewidth of an arbitrary KPZ system, having just fitted once and for all the constant $\gamma_1$.


\begin{figure}[t]
\centering
\includegraphics[width=0.48\textwidth]{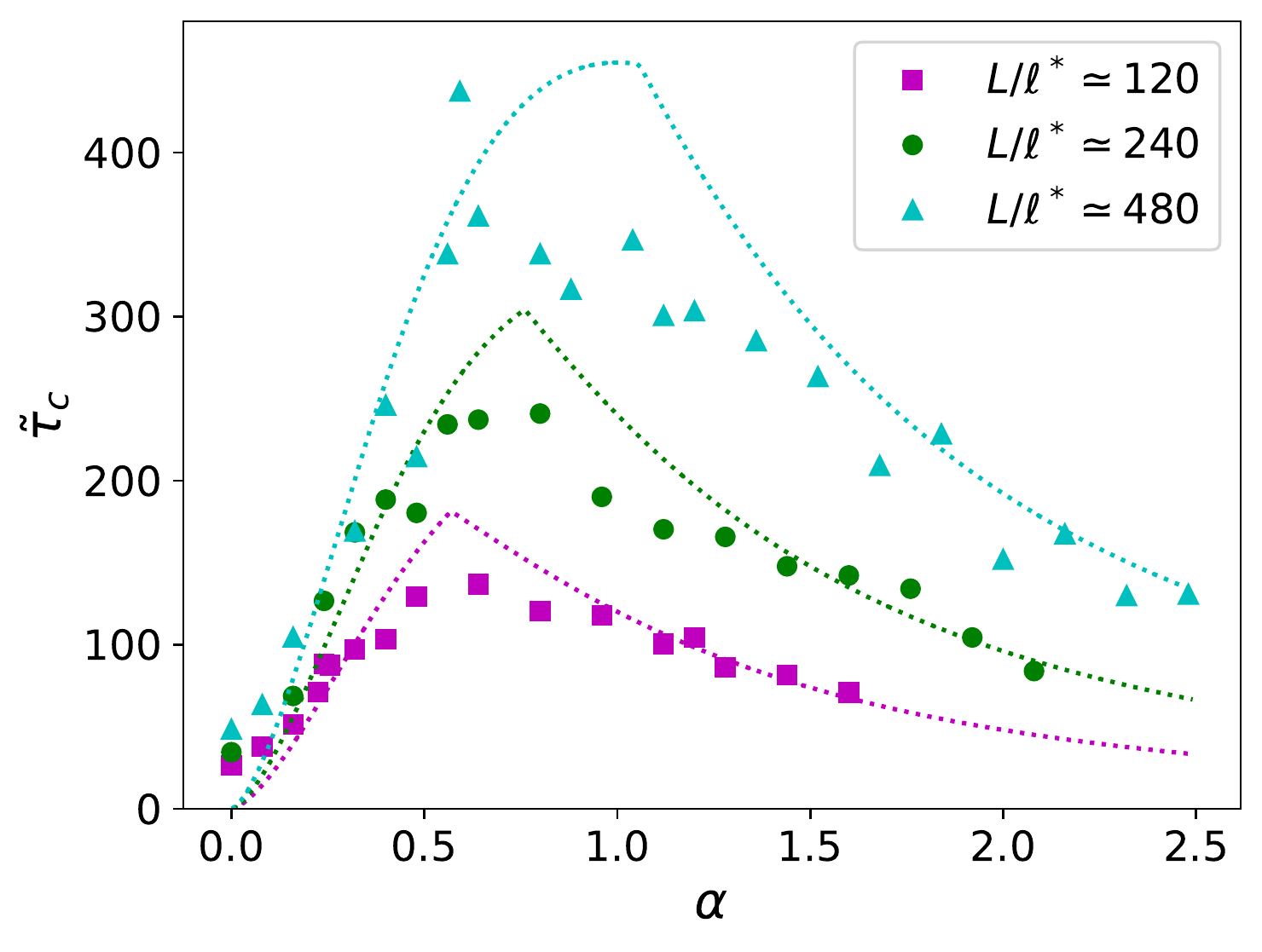}  
\caption{  Coherence time in natural units $\tilde{\tau}_c = (\phi^*)^2 \tau_c /t^*$ as a function of the Henry parameter
$\alpha=2gn_0/\Gamma$ for three different values of the system size.
The numerical points are obtained from simulations of the CGLE (\ref{eq:GPE}) using the same parameters as in Fig.~\ref{fig:Lscaling} except for $n_S=400$.
The dotted lines are the theoretical predictions: for large $\alpha$, we used the Henry-Schawlow-Townes formula
Eq.~(\ref{eq:Schawlow--Townes}); for smaller but non-zero $\alpha$, we used the KPZ scaling
Eq.~(\ref{eq:alfa_scaling}).
}
\label{fig:alfa_scaling}
\end{figure}

\subsection{Effect of finite interactions $g\neq 0$}

Let us know investigate the effect of a finite interaction constant, $g \neq 0$. 
On the one hand, the effective noise on the phase is enhanced via the same mechanism underlying the Henry broadening of the linewidth in the single-mode laser, as expressed by $\alpha$ in Eq.~(\ref{eq:Schawlow--Townes}).
On the other hand, for $g \neq 0$ the Laplacian term in the microscopic phase equation~(\ref{eq:KSE_tilde}) is nonzero, which tends to stabilize the fluid phase by reducing long wavelength fluctuations. As a consequence, the Bogoliubov-Gaussian theory holds up to larger system sizes and longer systems are needed to observe a clean KPZ scaling~\cite{gladilin2014}. 

The latter observation entails that the scaling of the linewidth with $g$ (or, more conveniently, with the adimensional $\alpha=2gn_0/\Gamma$ parameter) in a long system may be different from what the standard Henry broadening Eq.~(\ref{eq:Schawlow--Townes}) of short systems.
In the regime where the Laplacian term in KSE equation ~(\ref{eq:KSE_tilde}) dominates over the quartic derivative term, one can in fact approximate the KSE with a KPZ with $\nu \propto \alpha$ and $D \propto 1 + \alpha^2$.
The scaling with  $\alpha$ can then be straightforwardly obtained using formula (\ref{eq:KPZ_linewidth}), which yields
\begin{equation}
    \tilde{\tau}_c 
    =2/\gamma_{KPZ}(
    \tilde{\mu}, 
    2,
    1+\alpha^2,L/l^*)
    \propto 
    \frac{\alpha^{3/2}}{ (1 + \alpha^2)^{3/2}} \sqrt{{L}} .
    \label{eq:alfa_scaling}
\end{equation}
This scaling of the coherence time with $\alpha$ is reproduced by numerical simulations of the CGLE for three different sizes, as illustrated in Fig.~\ref{fig:alfa_scaling}. In these simulations, a smaller value of $n_S$ is used to increase fluctuations and observe the physics of interest within a feasible integration box. 

In order to highlight the general trends, the coherence time is plotted in natural units $\tilde{\tau}_c = {(\phi^*)^2}\tau_c/{t^*}$.
At very small $\alpha$, the
same behavior studied in Fig.~\ref{fig:Lscaling} for the $g=0$ case is recovered: in this case, the phase dynamics is described by a KSE with negligible Laplacian term 
and Eq.~(\ref{eq:alfa_scaling}) does not apply. As a result, the coherence time tends to a finite value in the $\alpha\to 0$ limit.

At small but finite values of $\alpha$, the Laplacian term starts dominating over the quartic term and the points follow the trend predicted by Eq.~(\ref{eq:alfa_scaling}).
Finally, at larger interactions, the spatial fluctuations are dominated by the strong Laplacian term, so the physics turns out to be well described again by Bogoliubov theory and the linewidth recovers the Henry scaling Eq.~(\ref{eq:Schawlow--Townes}).
The dotted lines represent the minimum between the coherence time predicted by 
Eq.~(\ref{eq:alfa_scaling}) and Eq.~(\ref{eq:Schawlow--Townes}) and the cusp that separating the two regimes is smoothened out by the competition between the different effects.

It is interesting to note how the window in which KPZ physics is observed gets wider in larger systems. In Fig.~\ref{fig:alfa_scaling}, this is visible as a shift of the cusp towards larger $\alpha$ for growing $L/\ell^*$.
Another interesting feature  visible in this plot is that for a given finite size, the optimal coherence time is achieved for intermediate values of the interaction constant $g$.

\section{Skewness and cumulant approximation}
\label{sec:skewness}

\begin{figure}[t]
\centering
\includegraphics[width=0.49\textwidth]{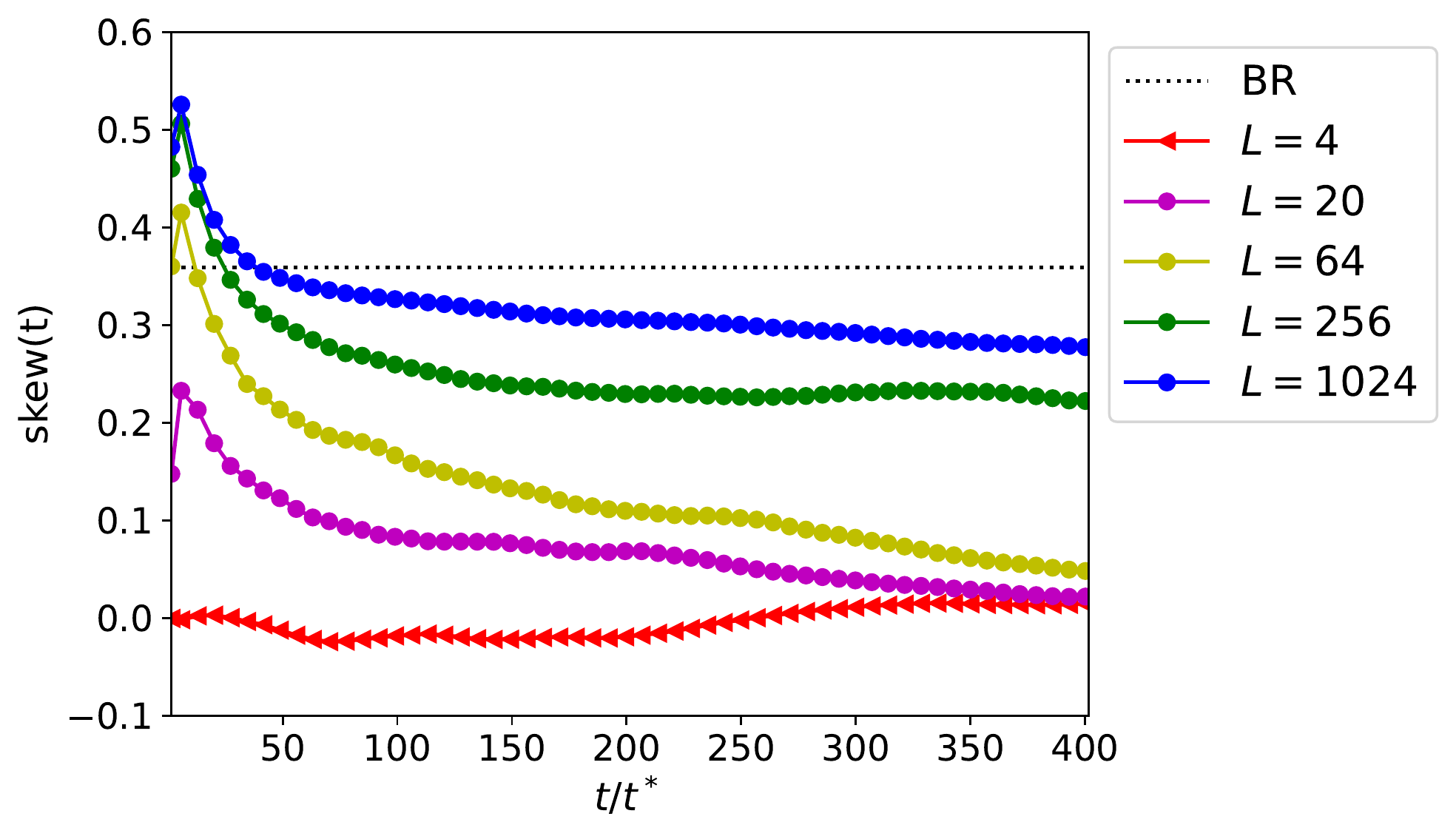}  
\caption{  Temporal evolution of the skewness $skew(t)$
of the phase for different system sizes, from the CGLE with the same parameters as in Fig.~\ref{fig:Lscaling}.
The dotted line is the value expected from the Baik-Rains distribution. 
}
\label{fig:skew}
\end{figure}

The analysis reported in the previous Sections heavily relies on the scaling behavior of the phase-phase correlator and the results have been translated to the $g^{(1)}$ under the so called cumulant approximation mentioned in Eq.~(\ref{eq:cumulant}).
More explicitly, this approximation can be formulated as
\begin{equation}
    \left|  \langle e^{-i\phi(x, t)} e^{i\phi(x, 0)} \rangle  \right|
\simeq
e^{-\frac{1}{2}
\Delta{\phi}_{{x},{t}}^2
},
\end{equation}
which is actually exact if 
$\phi(x, t)$ is a Gaussian random variable and is a good approximation if the field $\phi(x, t)$ is small: in this regime, one can in fact expand the exponential to second order in $\phi$, compute the phase-phase correlator (also called the second cumulant), and then re-exponentiate the result. 

However, in the KPZ regime the phase is not a Gaussian random variable and has a more complex statistics given in Eq.~(\ref{eq:phaseRV}). For a limited temporal window, the fluctuations of $\phi$ are small and one can recover the KPZ phase-phase correlator by taking the logarithm of $g^{(1)}(x,t)$~\cite{amelio2020theory}. At longer times, however, this procedure is no longer legitimate~\cite{fontaine2022}.

It is then even more remarkable to notice the excellent agreement between the CGLE and KSE predictions for the coherence time that is visible in Fig.~\ref{fig:Lscaling}. For this, we recall that the fitted quantity in the first case is the logarithm of $g^{(1)}(t)$, while it is directly the phase-phase correlator in the second case.

This result suggests that the finite system version of Eq.~(\ref{eq:phaseRV})
should have the form
\begin{equation}
\phi(x,t) - \phi(x,0) 
= v_{\infty} t + \sigma t^{1/3} X +  (\gamma_c t)^{1/2} Y + ...
\label{eq:phaseRVpro}
\end{equation}
where $X$ is a Baik-Rains-distributed random variable and $Y$ is a Gaussian variable. 
According to this formula, the long time decay of the coherence function is proportional to $\propto \exp\{-\frac{\gamma_c}{2} t + o(t) ... \}$, where the leading order term  is determined by the cumulant of the Gaussian term of Eq.~(\ref{eq:phaseRVpro}) and is safely obtained by exponentiation;
calculation of the subleading terms denoted as $o(t)$ would instead require a very nontrivial re-exponentiation of the stochastic variable in Eq.~(\ref{eq:phaseRVpro}).
When the logarithm of the numerically calculated $g^{(1)}(t)$ is fitted to extract the linewidth, no signatures of a deviation from linear behavior are found: most likely detection of the subleading terms would require extremely clean data on a very broad temporal window, which goes beyond our numerical possibilities.

Direct numerical evidence in support of Eq.~(\ref{eq:phaseRVpro}) is displayed in Fig.~\ref{fig:skew}, where we directly measured  the phase of the 
$g=0$
CGLE and computed the temporal evolution of its skewness  for different system sizes.
As usual, the skewness is defined as the normalized third cumulant
\begin{equation}
    skew(t) = \frac{\langle 
    (\phi(x,t) - \phi(x,0))^3
    \rangle}{
    \left[ \langle 
    (\phi(x,t) - \phi(x,0))^2
    \rangle \right]^{3/2}
    }
\end{equation}
and  is a measure of the asymmetry of a distribution. Within the temporal window  corresponding to the KPZ regime, the skewness is finite and has a magnitude comparable to the one expected from the Baik-Rains distribution. As expected, it gets smaller at later times, the crossover time depending on the system size. For system sizes that are too small to support the KPZ regime, the skewness remains always small.

Similarly to previous studies dealing with the KPZ sub-universality classes in polariton systems~\cite{squizzato2018}, the numerical load of this calculation makes it difficult to obtain a clean measurement of the skewness. Nevertheless, the available data confirm the excellent agreement between the linewidth obtained from $\log g^{(1)}(t)$ and $\Delta{\phi}_{{x},{t}}^2$ and further justify re-exponentiation at very long times.

\begin{figure}[t]
\centering
\includegraphics[width=0.46\textwidth]{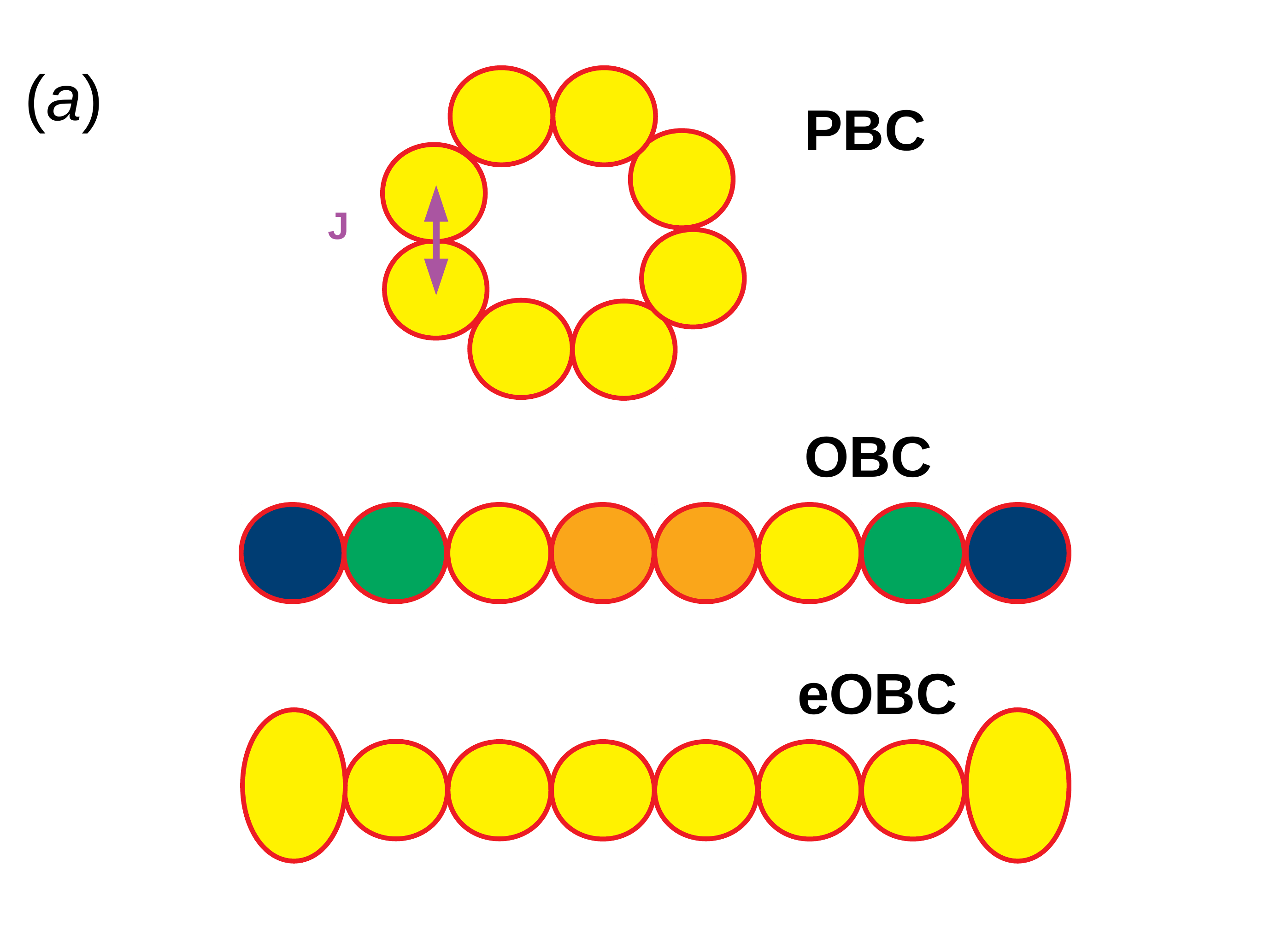}  
\includegraphics[width=0.48\textwidth]{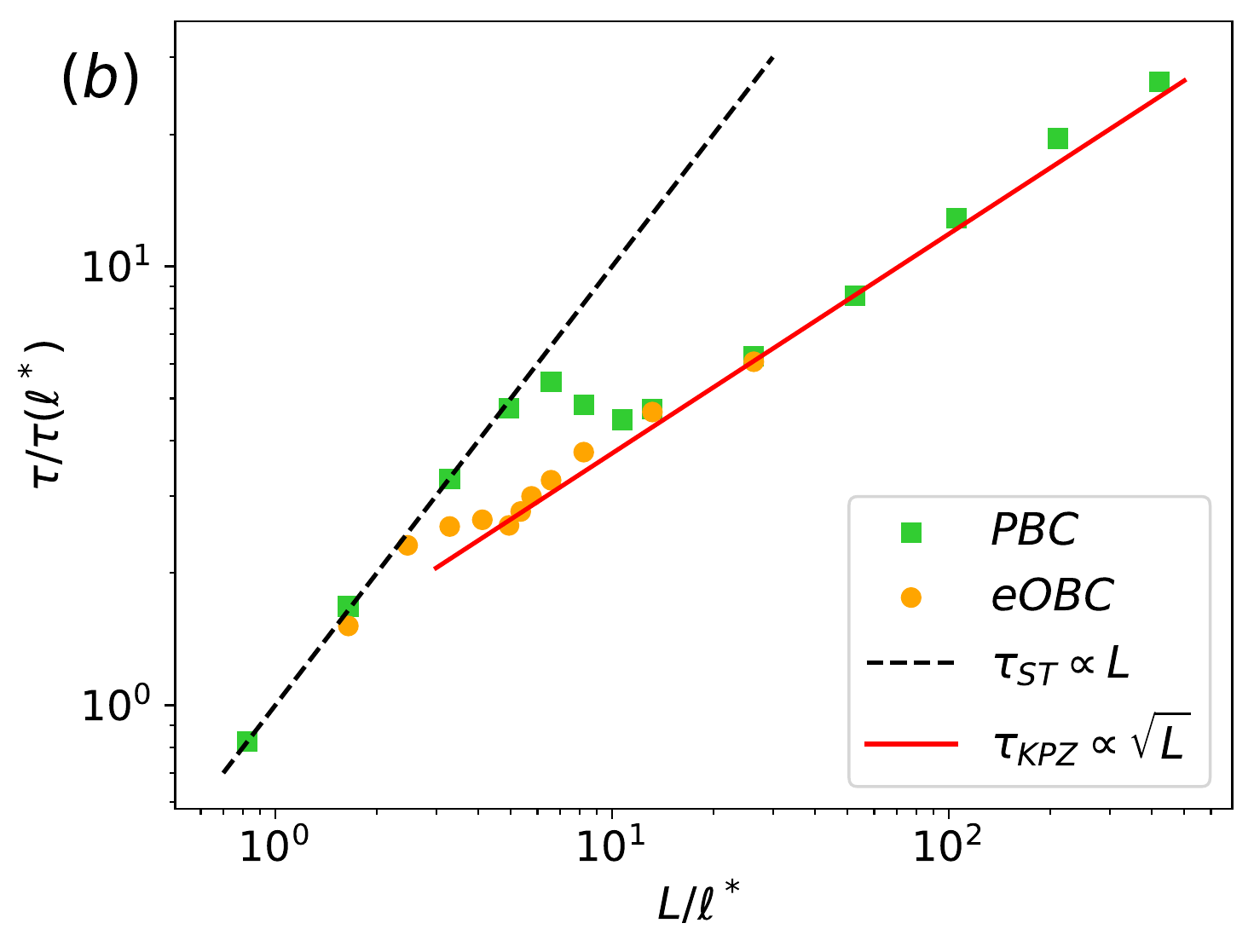}  
\caption{
(a) Sketch of different boundary conditions for a discrete lattice geometry. The color code indicates the spatial intensity profile of the steady state condensate: the enhanced open boundary conditions (eOBC) introduced in Eq.~(\ref{eq:eOBC}) allow to obtain a uniform density.
(b) Coherence time as a function of the system size in the two cases of PBC (green) and eOBC (orange), plotted in the same units as in Fig.~\ref{fig:Lscaling}. The system with eOBC supports Bogoliubov modes with longer lifetime and, thus, displays an earlier departure from the Schawlow-Townes scaling at smaller system sizes.
}
\label{fig:obc}
\end{figure}

\section{The role of boundary conditions}
\label{sec:boundaryconditions}

A crucial issue in view of experiments is to understand how the boundary conditions affect the dynamics of fluctuations and, then, the linewidth. 
To highlight the analogy with the most promising device used in~\cite{fontaine2022} and, at the same time, avoid UV regularizations,
in this Section we focus on the case of a discrete lattice of resonators 
with hopping $J$, whose single-particle conservative Hamiltonian will be denoted $H_{latt}$.
So far we were concerned with systems with periodic boundary conditions (PBC), as in the top sketch of Fig.~\ref{fig:obc}(a). While experimental implementation of a 1D device with periodic boundary conditions is possible, it may be in practice not straightforward~\cite{Contractor2022}.

Standard systems realize in fact open boundary conditions (OBC). From the point of view of critical systems, this configuration presents serious drawbacks, since the uniform state is no longer an eigenstate of $H_{latt}^{OBC}$. This leads to a non-uniform spatial shape of the condensate mode, involving reflection from the two endpoints and complex interference phenomena, as shown in the central sketch of Fig.~\ref{fig:obc}(a).

A practical way around, adopted in several current experiments, is to restrict pumping to the central part of a very long lattice. In spite of the ensuing outward current~\cite{wouters2008}, this configuration allowed to observe KPZ universality~\cite{fontaine2022} and address the linewidth problem. While flow can be avoided by imposing an additional harmonic confinement, as numerically considered in~\cite{deligiannis2020}, quantitative studies of finite size effects would still benefit from using a spatially uniform condensate and exploting the full length of the avaiable device.

In what follows,  we propose enhanced open boundary conditions (eOBC) which allow for a uniform condensate in a spatially finite system. In this scheme, the energies of the two extremal resonators are lowered by an amount $J$, so the Hamiltonian for the $N$-site lattice system reads
\begin{equation}
    H_{latt}^{eOBC} =
    -J \sum_{i=2}^N \left[
     \psi^{\dagger}_i \psi_{i-1} +
    \psi^{\dagger}_{i-1} \psi_{i}
    \right]
    - J \left[
    \psi^{\dagger}_1 \psi_1 +
    \psi^{\dagger}_N \psi_N
    \right].
    \label{eq:eOBC}
\end{equation}
In such a configuration, it is straightforward to verify that the uniform state is an eigenvector $H_{latt}^{eOBC}$ and a steady-state for the lasing system, as sketched in the last row of Fig.~\ref{fig:obc}(a). We expect that the required design of the endpoint resonators can be straightforwardly realized in polariton micropillar systems.

Let us now consider the dynamics of fluctuations in these eOBC systems.
As it can be checked from numerical diagonalization, 
the second slowest Bogoliubov mode after the Goldstone consists of a cosine-like wavefunction of wavelength $2L$. While such a wavelength would not fit in a PBC system, it is allowed in our eOBC lattice thanks to the relaxed wavelength quantization constraint. 
As a consequence, in eOBC one requires half the length $L$ to display an equally long-lived mode as in PBC. In particular, this reduction affects the critical size at which the linewidth departs from the Bogoliubov-Schawlow-Townes prediction and starts showing KPZ features. In Fig.~\ref{fig:obc}(b), the normalized coherence time is plotted as a function of the system length in natural units. Here, the prediction for the CGLE in PBC already shown in Fig.~\ref{fig:Lscaling} (green points) is compared to the one of the CGLE with eOBC (orange points): in agreement with our expectations, the Bogoliubov-Schawlow-Townes theory breaks down at a smaller size $L/\ell^* \sim 2.5$ for eOBC compared to the size $L/\ell^* \sim 5$ for PBC. Beyond the crossover, KPZ physics sets in in both cases and the universal properties are the same.

\section{Conclusions}
\label{sec:conclusions}

In this work, we have studied the long-time decay of the emission coherence of a one-dimensional non-equilibrium condensate. An exponential decay always overtake at long enough times in finite-length systems, so that the emission linewidth can be seen as a finite size effect. Depending on the system length, two regimes can be identified: for short systems, one can apply a linearized Bogoliubov theory and find the usual linear scaling of the coherence time with the system size originally predicted by Schawlow and Townes. On the other hand, for long wires the scaling of the linewidth is dominated by effects beyond Bogoliubov and displays a Kardar-Parisi-Zhang critical behaviour, leading to a square-root dependence on the length.

Markedly different roles of the optical nonlinearities on the linewidth are highlighed. On the one hand, optical nonlinearities increase the damping rate of the Bogoliubov modes belonging to the diffusive Goldstone branch and correspondingly reinforce the Laplacian term of the KPZ equation: this enhances the spatial stiffness of the phase dynamics and tends to prolonge the coherence time. On the other hand, the same optical nonlinearities are responsible for a Henry broadening effect, which reinforces noise and thus tends to reduce the coherence time.
The interplay of these effects leads to a very nontrivial scaling of the linewidth with the optical nonlinearity strength and the system size, the optimal coherence being achieved for intermediate values of the nonlinearity.

 This behaviour was demonstrated numerically by solving the full field equation in the form of a stochastic Complex Ginzburg Landau Equation, and then explained in terms of a finite-size scaling hypothesis. Successful comparison of our results with a simulation of the Kuratomoto-Sivashinsky equation confirms that our conclusion are due to the phase dynamics. We then show how the linewidth extracted the logarithm of $g^{(1)}(t)$ matches with the diffusion rate in the phase-phase correlator. This suggests that the cumulant approximation is legitimate, at least at large times and is explained by monitoring the temporal evolution of the skewness of the phase, which is shown to decay in time. On this basis, we conjecture that, due to the finite size of the system, the phase distribution shifts from a Baik-Rains to a Gaussian form at very long time, which justifies the accuracy of the re-exponentiation procedure. 

We finally discuss the different experimental platforms where our predictions can be investigated. In particular, lattice geometries with enhanced open boundary conditions are proposed, which support a spatially uniform steady-state lasing mode and thus facilitate experimental investigations of the scaling with the system size.

Open theoretical questions include a full understanding of the non-monotonic feature visible in Fig.~\ref{fig:Lscaling} at the crossover between the Schawlow-Townes and KPZ scaling, a general study of the scaling of the linewidth with system size in higher dimensions and the impact of multi-mode laser operation on the linewidth.

\section*{Acknowledgements}
We are grateful to Leonie Canet, Erwin Frey, Anna Minguzzi and Davide Squizzato for useful discussions.
We acknowledge financial support from \iac{the H2020-FETFLAG-2018-2020 project "PhoQuS" (n.820392),} from the Provincia Autonoma di Trento, and from the PNRR MUR project PE0000023-NQSTI. All numerical calculations were performed using the Julia Programming Language \cite{bezanson2017}.

\appendix
\section{Bogoliubov modes}

\begin{figure}[h]
\centering
\includegraphics[width=0.48\textwidth]{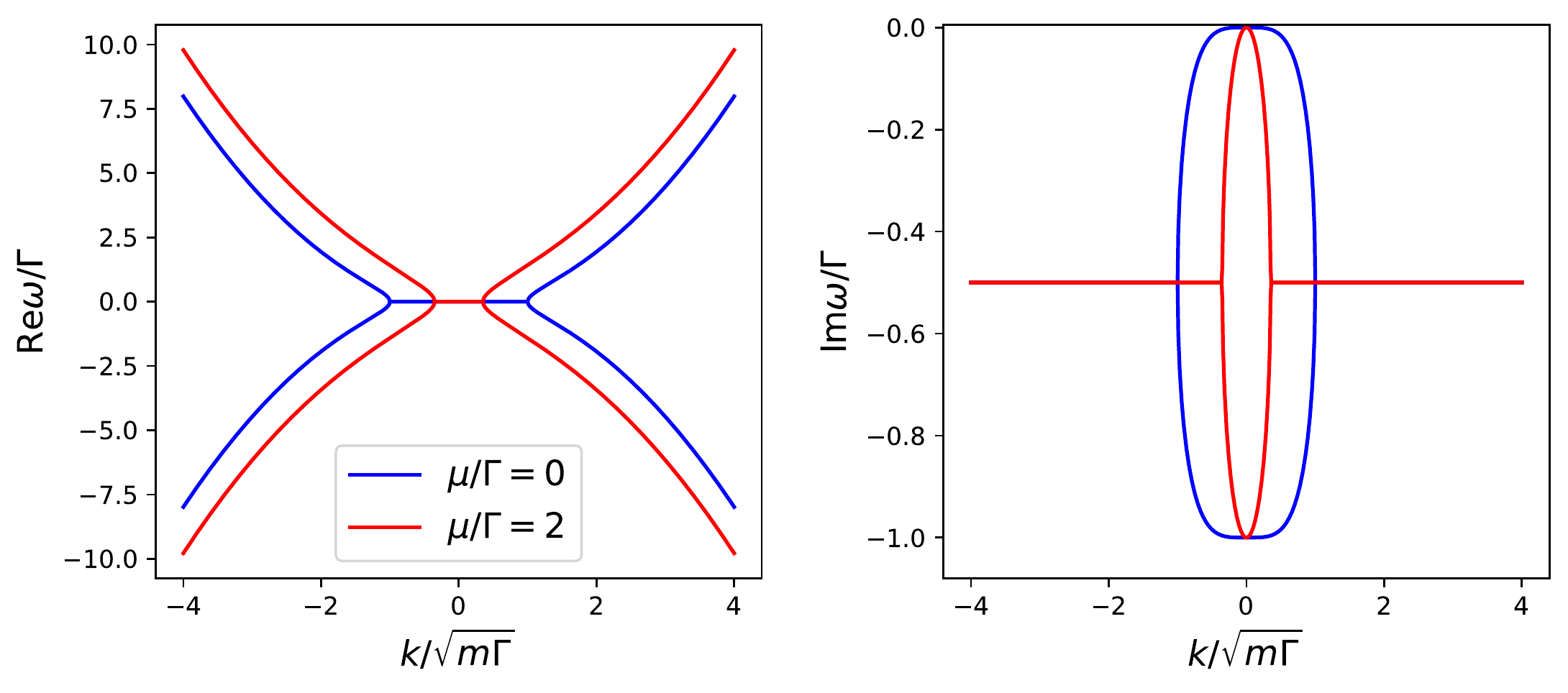}  
\caption{Real (left) and imaginary (right) part of the Bogoliubov dispersion of collective excitations Eq.~(\ref{eq:bogo_modes}) as a function of momentum $k$. Blue (red) points refer to vanishing (finite) values of the optical nonlinearity $g$. 
}
\label{fig:bogo}
\end{figure}

In this Appendix, we briefly review the Bogloliubov theory of the collective excitations on top of a non-equilibrium condensate~\cite{wouters2007,chiocchetta2013}. The linearized perturbations on top of the spatially uniform mean-field steady-state $\psi=\sqrt{n_0}$ have 
a complex frequency dispersion
\begin{equation}
\omega_{\pm}(k)
= -i \frac{\Gamma}{2} \pm
\sqrt{\frac{k^2}{2m} \left(
\frac{k^2}{2m} + 2 \mu \right)
- \left(\frac{\Gamma}{2}\right)^2 + i0^+}
\label{eq:bogo_modes}
\end{equation}
as a function of momentum $k$.  
The real and imaginary parts of the Bogoliubov spectrum are displayed in the left and right panels of Fig.~\ref{fig:bogo}, respectively.
For stronger optical nonlinearities $\mu=g\,n_0$, the size of the diffusive region shrinks in $k$.

The soft mode with $\omega_+(k=0) = 0$ is the Goldstone mode associated to the spontaneous symmetry breaking mechanism that fixes the condensate phase.
The low-$k$ long-wavelength region  of the  Goldstone branch consists of phase-like modes and recovers the linear part of the KSE (\ref{eq:KSE}), with the usual Laplacian and quartic derivative terms. 
The amplitude mode corresponds instead to density fluctuations and shows a finite damping rate $\Gamma$ in the $k\to 0$ long-wavelength limit.
The amplitude and Goldstone branches merge at larger momenta to yield single-particle-like modes with a parabolic dispersion and a constant decay rate $\Gamma/2$. 

As long as the Bogoliubov theory holds, only the  free diffusion of the Goldstone mode contributes to the linewidth. All other modes have in fact a finite lifetime and do not play any significant role at very long times~\cite{amelio2020theory}. On this basis, we refer to the Schawlow-Townes linewidth (that only involves the single condensed mode) as the Bogoliubov linewidth~~\cite{amelio2022bogoliubov}.
A geometric explanation of the Henry broadening is that the nonlinearity $\alpha$  makes the the Goldstone and amplitude modes non-orthogonal~\cite{amelio2022bogoliubov}.






\bibliography{bibliography}

\end{document}